\documentclass[english,aps,prl,reprint]{revtex4-1}
\usepackage[T1]{fontenc}
\usepackage[utf8]{inputenc}
\usepackage{float}
\usepackage{amsmath}
\usepackage{amssymb}
\usepackage{graphicx}

\makeatletter
\@ifundefined{showcaptionsetup}{}{%
 \PassOptionsToPackage{caption=false}{subfig}}
\usepackage{subfig}
\makeatother

\usepackage{babel}
\begin{document}
\global\long\def\hc{\text{h.c.}}
\global\long\def\re{\text{Re}}
\global\long\def\im{\text{Im}}
\global\long\def\K{\mbox{ K}}
\global\long\def\meV{\mbox{ meV}}
\global\long\def\eV{\mbox{ eV}}
\global\long\def\hf{\!\,_{2}F_{1}}
\global\long\def\pr{\prime}
\global\long\def\prpr{\prime\prime}
\global\long\def\sgn{\text{sgn}}

\title{Scale-invariance as the cause of the superconducting dome in the cuprates}

\author{Zhidong Leong}

\author{Kridsanaphong Limtragool}

\author{Chandan Setty}

\author{Philip W. Phillips}

\affiliation{Department of Physics and Institute for Condensed Matter Theory,
University of Illinois, Urbana, Illinois 61801, U.S.A}

\date{\today}
\begin{abstract}
Recent photoemission spectroscopy measurements (T. J. Reber et al., arXiv:1509.01611) of cuprate superconductors have inferred that the self-energy exhibits critical scaling over an extended doping regime, thereby calling into question the conventional wisdom that critical scaling exists only at isolated points.  In particular, this new state of matter, dubbed a power-law liquid, has a self-energy whose imaginary part scales as $\Sigma^{\pr\pr}\sim(\omega^{2}+\pi^{2}T^{2})^{\alpha}$, with $\alpha=1$ in the overdoped Fermi-liquid state and $\alpha\leq0.5$ in the optimal to underdoped regime. Previously, we showed that this self-energy can arise from interactions between electrons and unparticles, a scale-invariant sector that naturally emerges from strong correlations. Here, taking the self-energy as a given, we first reconstruct the real part of the self-energy. We find that the resultant quasiparticle weight vanishes for any doping level less than optimal, implying an absence of particle-like excitations in the underdoped regime.  Consequently, the Fermi velocity vanishes and the effective mass diverges for $\alpha\leq\frac{1}{2}$, in agreement with earlier experimental observations.  We then use the self-energy to reconstruct the spectral function and compute the superconducting $T_c$ within the BCS formalism. We find that the $T_c$ has a dome-like structure, implying that broad scale invariance manifested in the form of a power-law liquid is the likely cause of the superconducting dome in the cuprates.  
\end{abstract}
\maketitle

\section{Introduction}

Understanding the physics of cuprate superconductors involves identifying
the low-energy degrees of freedom responsible for the normal state's
anomalous features, such as $T$-linear resistivity, pseudogap, and
Fermi arcs. In general, the electron Green function can be written
as $G\left(k,\omega\right)=\left[\omega-\epsilon_{k}-\Sigma\left(\omega\right)\right]^{-1}$,
where $\epsilon_{k}$ is the bare energy spectrum, and $\Sigma$ is
the self-energy. Recent angle-resolved photoemission spectroscopy
(ARPES) measurements \cite{Reber2015} of the cuprates have revealed
that the imaginary part of the electron self-energy has the scaling
form

\begin{eqnarray}
-\Sigma^{\prpr}\left(\omega\right) & = & \Gamma_{0}+\lambda\frac{\left(\omega^{2}+\pi^{2}T^{2}\right)^{\alpha}}{\omega_{N}^{2\alpha-1}},\label{eq:PLL-imSigma}
\end{eqnarray}
over a wide range of doping. The key parameter here is the scaling
exponent $\alpha$, which varies from $\alpha=1$ in the overdoped
Fermi-liquid state to $\alpha=\frac{1}{2}$ at optimal doping, and
to $\alpha<\frac{1}{2}$ at underdoping. Other relevant parameters
include a dimensionless coupling constant $\lambda\sim0.5$, a high-energy
scale $\omega_{N}\sim0.5\eV$ to maintain dimensional consistency,
and an impurity scattering term $\Gamma_{0}\sim8\text{ to }35\meV$.

What is new here is that this scaling form persists over a wide range of doping, manifesting not just at a single point as traditional critical scenarios would suggest. Given the novelty of this scaling form, it is peculiar that the full consequences of this power-law scaling have not been explored previously. It is just this task that
we perform here. We explore the consequences for 1) the Fermi velocity, 2) the effective mass, 3) the quasiparticle weight, and 4) the superconducting dome.  All these quantities reveal truly unusual behaviors that are directly related to the power-law liquid's unconventional scaling observed in the experiments.   

Theoretically, mechanisms yielding non-Fermi-liquid scalings have
been extensively studied \cite{Varma1989,Metzner2003,Lee2009a,Watanabe2014a,Fitzpatrick2014,Sachdev2011,Faulkner2010a,Casey2011,Faulkner2010b,Zaanen2015}.
In a marginal Fermi liquid \cite{Varma1989}, a polarizability proportional
to $\omega/T$ leads to $T$-linear resistivity, while a $d$-wave
Pomeranchuk instability in two dimensions \cite{Metzner2003} yields
self-energies with $\omega^{2/3}$ and $T^{2/3}$ dependence. In addition,
similar behaviors can also be obtained by coupling quasiparticles
with gauge bosons \cite{Lee2009a}, Goldstone bosons \cite{Watanabe2014a},
and critical bosons \cite{Fitzpatrick2014} near a quantum critical
point \cite{Sachdev2011}. Furthermore, strong coupling theories using
the anti-de Sitter spacetime (AdS)/conformal field theory (CFT) correspondence
\cite{Faulkner2010a} and Gutzwiller projection in hidden Fermi-liquid
theory \cite{Casey2011} also exhibit $T$-linear resistivity. In
particular, the spectral functions calculated within the AdS/CFT formalism
can also exhibit a range of power-law scaling when the scaling dimension
of the boundary fermionic operator is tuned continuously \cite{Faulkner2010b,Zaanen2015}. 

Given the interest in experimentally relevant self-energies for the cuprates, it is truly remarkable
that the experimental consequences of the power-law liquid have not been explored until now.  Specific to Eq. \ref{eq:PLL-imSigma}, since the scaling form is robust
up to $0.1\eV$ and $250\K$ \cite{Reber2015}, we showed previously
that such a behavior can originate from interactions between electrons
and unparticles, a scale-invariant sector that naturally emerges due
to strong correlations in the cuprates \cite{Limtragool2016,Leong2017}.
Originally proposed as a scale-invariant sector within the standard model \cite{Georgi2007}, unparticles can arise in the cuprates because any nontrivial infrared dynamics in a strongly correlated electron system is controlled by a critical fixed point. Consequently, scale invariance can be used to construct the form of the underlying propagator. This propagator which can acquire an anomalous dimension within the renormalization group approach is the unparticle propagator. Furthermore, in the context of AdS/CFT, one of us \cite{LaNave2016,LaNave2017} showed that a massive scalar field in the bulk is generally dual to a nonlocal operator (i.e., a fractional Laplacian) on the boundary. The propagator of these operators is of a power-law form, just like the unparticle propagator. These results indicate that unparticles should generically exist in a strongly coupled system. 

In the context of the cuprates, unparticles have been proposed to explain the absence of Luttinger's theorem in the pseudogap phase \cite{Phillips2013} using zeros in the Green function \cite{Dave2013} and have also been found to yield unusual superconducting properties \cite{Phillips2013,LeBlanc2015,Karch2016} and optical conductivity \cite{Limtragool2015}. 

In particular, a power-law liquid can be obtained from interactions between electrons and unparticles \cite{Limtragool2016,Leong2017}. The propagator of fermionic unparticles can be written as $G_u\left(k,i\omega_n\right) = [i\omega_n-\epsilon^u_k]^{-1+d_u}$, where $d_u$ is the anomalous dimension and $\epsilon^u_k$ is the energy spectrum of unparticles. Due to the branch cut in the unparticle propagator, the scattering phase space for electron-unparticle interactions is nontrivially altered. Consequently, the electron self-energy due to such interactions scales with energy and temperature, with the scaling exponent $\alpha$ dependent on the anomalous dimension $d_u$ of the unparticle propagator as $d_u=\alpha-1$ \cite{Leong2017}. 

In this paper, we study the superconducting $T_{c}$ of a power-law
liquid. Within the BCS formalism, we show that the $T_{c}$ is non-monotonic
with respect to $\alpha$, the self-energy scaling exponent. The $T_{c}$
peaks at $\alpha=\frac{1}{2}$, reproducing the cuprates' superconducting
dome. We attribute this behavior to the scaling form of the electron
spectral function at low energies, where the scaling exponent is minimum
at $\alpha=\frac{1}{2}$. Furthermore, we find that, due to strong
renormalization of the spectral weights towards the Fermi level, the
Fermi velocity vanishes and the effective mass diverges for $\alpha\leq\frac{1}{2}$,
in agreement with earlier experimental observations \cite{Vishik2010,Sebastian2010,Singleton2010}.
Our results suggest that a power-law liquid contains physics central
to understanding the cuprates.

\section{Normal state properties}

\begin{figure}[H]
\begin{centering}
\includegraphics[scale=0.5]{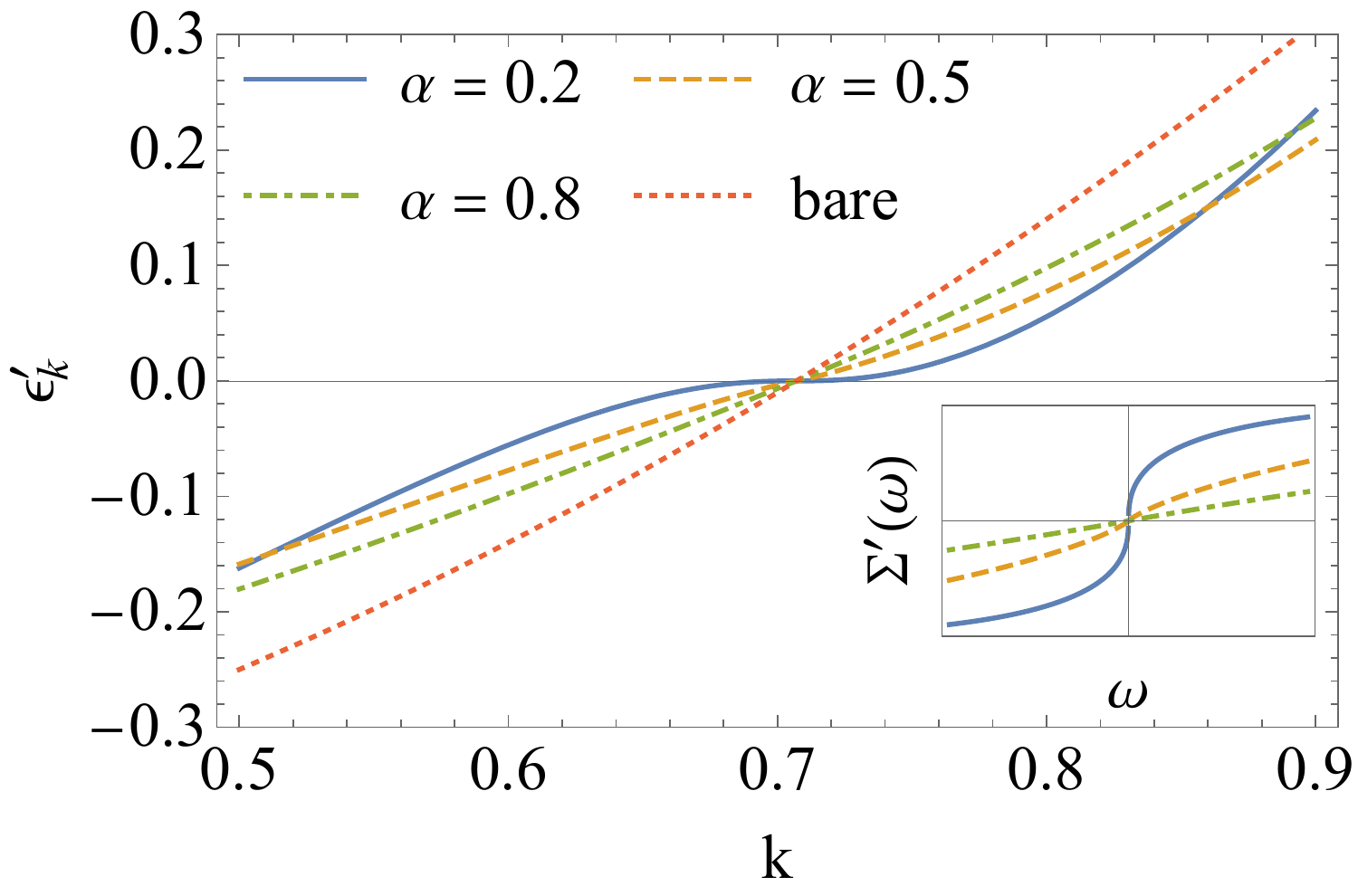}
\par\end{centering}
\caption{The energy spectrum of a power-law liquid is strongly renormalized
towards the Fermi level for $\alpha<\frac{1}{2}$, because the real
part of the self-energy $\Sigma^{\protect\pr}$ is non-analytic at
$\omega=0$, as shown in the inset. The impurity scattering term $\Gamma_{0}$
fixed at $0.01\omega_{N}$. \label{fig:energy-spectrum-vs-k}}
\end{figure}

The first obvious quantity to calculate is the real part of the electron self energy. 
This can be done directly from the Kramers-Kronig relationship: 
\begin{eqnarray}
\Sigma^{\pr}\left(\omega\right) & = & \frac{1}{\pi}\mathcal{P}\int d\omega^{\pr}\frac{\Sigma^{\prpr}\left(\omega^{\pr}\right)}{\omega^{\pr}-\omega}.
\end{eqnarray}
Interested in only the low energy behavior, we integrate up to the
high-energy scale $\omega_{N}$, assuming that the effects of $\Sigma^{\pr\pr}$
at higher energies are negligible. Also, for the integral to be analytically
tractable, we omit the $T$-dependence in the self-energy. As detailed
in App. \ref{sec:reSigma-derivation}, we obtain \begin{widetext}

\begin{eqnarray}
\Sigma^{\pr}\left(x\omega_{N}\right) & = & \frac{2\Gamma_{0}}{\pi}\text{artanh}x-\lambda\omega_{N}\tan\left(\alpha\pi\right)\text{sgn}x\left|x\right|^{2\alpha}-\frac{\lambda\omega_{N}}{2\alpha\pi}\left[\hf\left(1,-2\alpha;1-2\alpha;x\right)-\hf\left(1,-2\alpha,1-2\alpha,-x\right)\right],\label{eq:PLL-reSigma}
\end{eqnarray}

\end{widetext} where $\hf\left(a,b;c;z\right)$ is the hypergeometric
function. Illustrated in Fig. \ref{fig:energy-spectrum-vs-k} (inset),
this result strongly influences several low-energy behaviors of a
power-law liquid. For notational simplicity, we measure energies in units of $\omega_{N}$. For concreteness, we consider a quadratic bare energy
spectrum $\epsilon_{k}=k^2-k_F^2$ in two dimensions, with Fermi momentum $k_F=1/\sqrt{2}$.  Since our focus is
the $\alpha$-dependence of low-energy properties, we fix $\Gamma_{0}$
at a constant value of $0.01$.

The renormalized band $\epsilon_{k}^{\pr}$ is determined by
$\epsilon_{k}^{\pr}-\epsilon_{k}-\Sigma^{\pr}\left(\epsilon_{k}^{\pr}\right)=0$.
Fig. \ref{fig:energy-spectrum-vs-k} shows that, close to the Fermi
level, the bare dispersion is strongly renormalized towards the Fermi
level for $\alpha<\frac{1}{2}$. This is quantified by the Fermi velocity
$v_{F}$, which is renormalized by the quasiparticle residue $Z$
via $v_{F}^{\pr}=Zv_{F}$, where
\begin{eqnarray}
Z & = & \left(1-\left.\frac{d\Sigma^{\pr}}{d\omega}\right|_{\omega=0}\right)^{-1}\nonumber \\
 & = & \begin{cases}
\left[1-\frac{2\bar{\Gamma}_{0}}{\pi}+\frac{2\lambda}{\left(2\alpha-1\right)\pi}\right]^{-1}, & \alpha>\frac{1}{2},\\
0, & \alpha\leq\frac{1}{2}.
\end{cases}\label{eq:PLL-Z}
\end{eqnarray}
A similar result was obtained in Ref. \onlinecite{Reber2015}. The
two cases arise due to the $\left|x\right|^{2\alpha}$ term in $\Sigma^{\pr}$.
The quasiparticle residue $Z$ of a Fermi liquid quantifies how particle-like
the system is, with unity denoting completely particle-like. The vanishing
of the quasiparticle residue for $\alpha\leq\frac{1}{2}$ therefore
reflects the absence of any particle-like behavior in a power-law
liquid, indicative of the strong correlations in underdoped cuprates.
A similar behavior also exists in an ultracold Fermi gas with strong
interactions \cite{Sagi2015}.

\begin{figure}[H]
\begin{centering}
\includegraphics[scale=0.5]{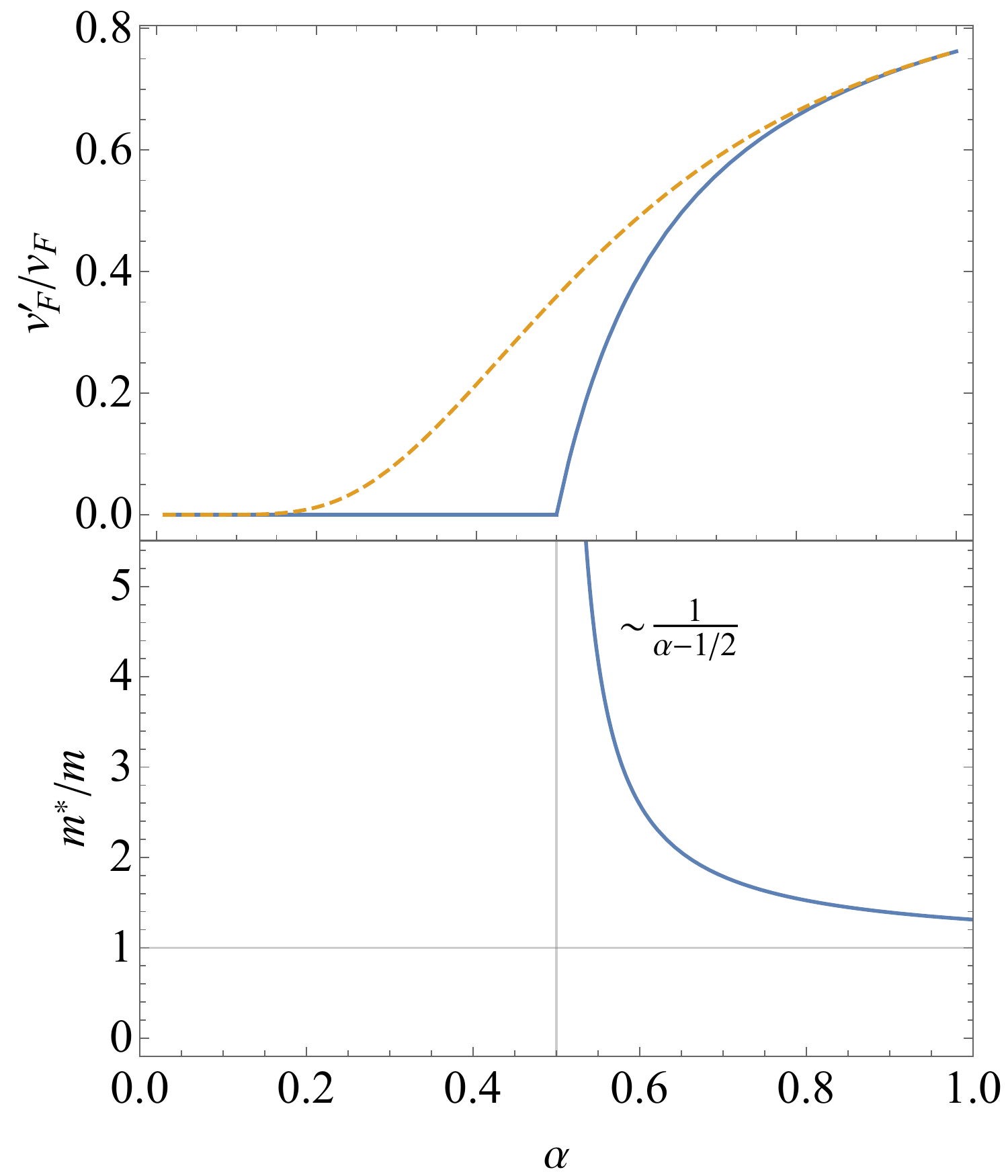}
\par\end{centering}
\caption{Top: (Solid) The Fermi velocity $v_{F}$ of a power-law liquid vanishes
for $\alpha\leq\frac{1}{2}$. (Dashed) The velocity just ($0.01\omega_{N}$)
below the Fermi level decreases monotonically with $\alpha$, in agreement
with ARPES measurements \cite{Vishik2010}. Bottom: The effective
mass of a power-law liquid diverges for $\alpha\leq\frac{1}{2}$,
in agreement with quantum oscillation measurements \cite{Sebastian2010,Singleton2010}.
\label{fig:vF-m-vs-alpha}}
\end{figure}

Shown in Fig. \ref{fig:vF-m-vs-alpha}, the Fermi velocity $v_{F}$
vanishing for $\alpha\leq\frac{1}{2}$ quantifies the strong renormalization
of the band towards the Fermi level. Experimentally, the Fermi velocity
can be determined from the slope of the band close to, but not exactly
at, the Fermi level. In Ref. \onlinecite{Vishik2010}, ARPES measurements
of the nodal Fermi velocity within $7\meV$ of the Fermi level show
that the Fermi velocity decreases monotonically with underdoping.
This behavior is reproduced in Fig. \ref{fig:vF-m-vs-alpha}, which
shows that the power-law liquid's velocity just below the Fermi level
decreases with $\alpha$. 

The vanishing Fermi velocity also implies that the effective mass
$m^{*}=k_{F}/v_{F}^{\pr}$ diverges as $\left(\alpha-\frac{1}{2}\right)^{-1}$,
as shown in Fig. \ref{fig:vF-m-vs-alpha}. This behavior has been
observed in the cuprates via quantum oscillations measurements \cite{Sebastian2010,Singleton2010}
and is attributed to a metal-insulator transition beneath the superconducting
dome. Our results thus far are robust in the sense that they are independent
of the values of $\Gamma_{0}$ and $\lambda$.

It is important to note that while ARPES measured the self-energy $\Sigma^{\prime\prime}$ over a limited energy range, the behavior of $\Sigma^{\prime\prime}$ at high energies is immaterial to our key result that the Fermi velocity $v_F$ vanishes when $\alpha\leq \frac{1}{2}$. This is because the vanishing of $v_F$ arises from the divergence of $d\Sigma^\prime/d\omega$ at $\omega=0$. From the form of the Kramers-Kronig relation, $\Sigma^{\prime\prime}$ at high energies has a finite contribution to $d\Sigma^\prime/d\omega|_{\omega=0}$ and so cannot affect the presence of the divergence.

\begin{figure}[H]
\subfloat[\label{fig:spectral-function-3D-plot}]{\begin{centering}
\includegraphics[scale=0.5]{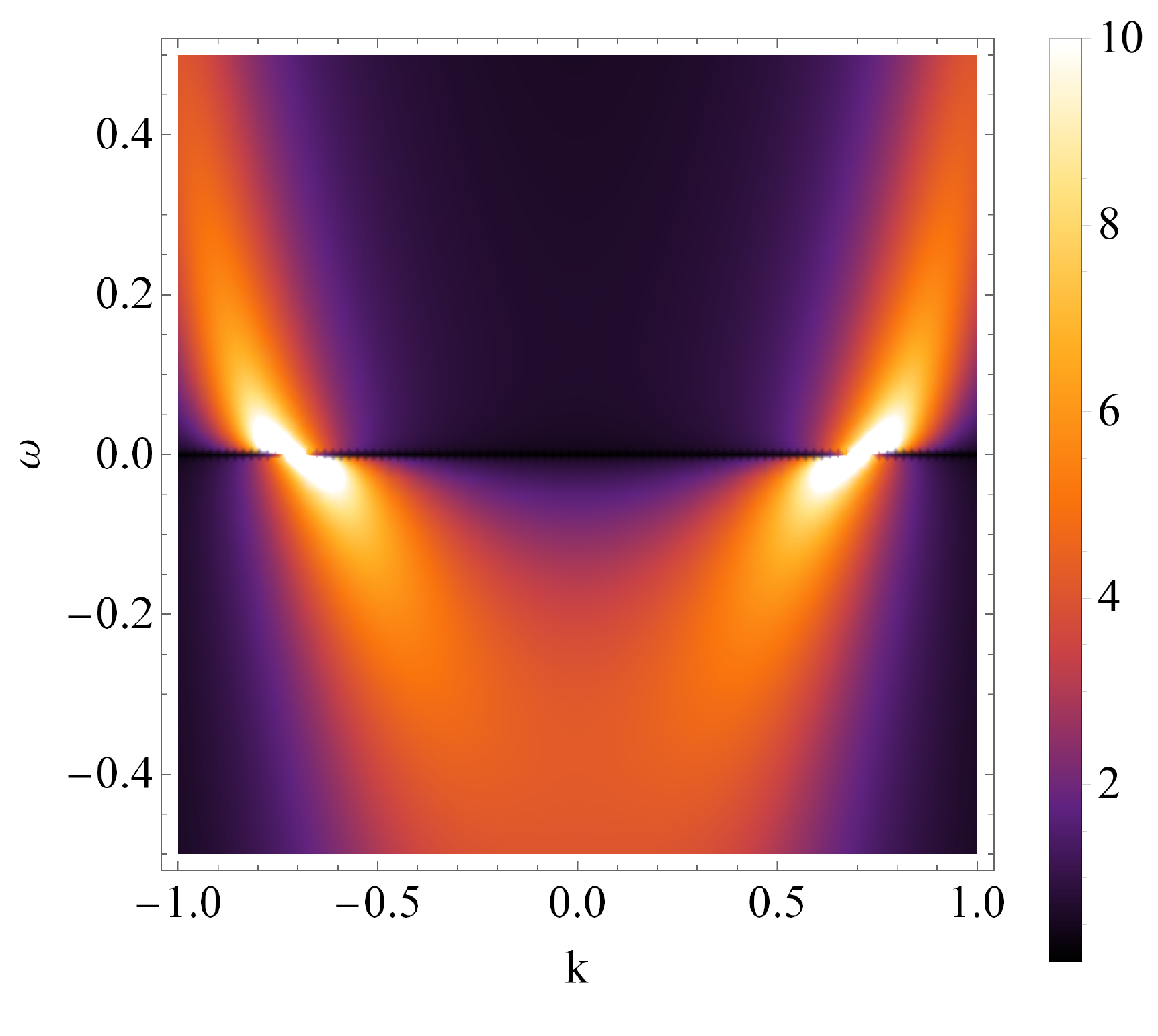}
\par\end{centering}
}

\subfloat[\label{fig:spectral-function-vs-omega} ]{\begin{centering}
\includegraphics[scale=0.5]{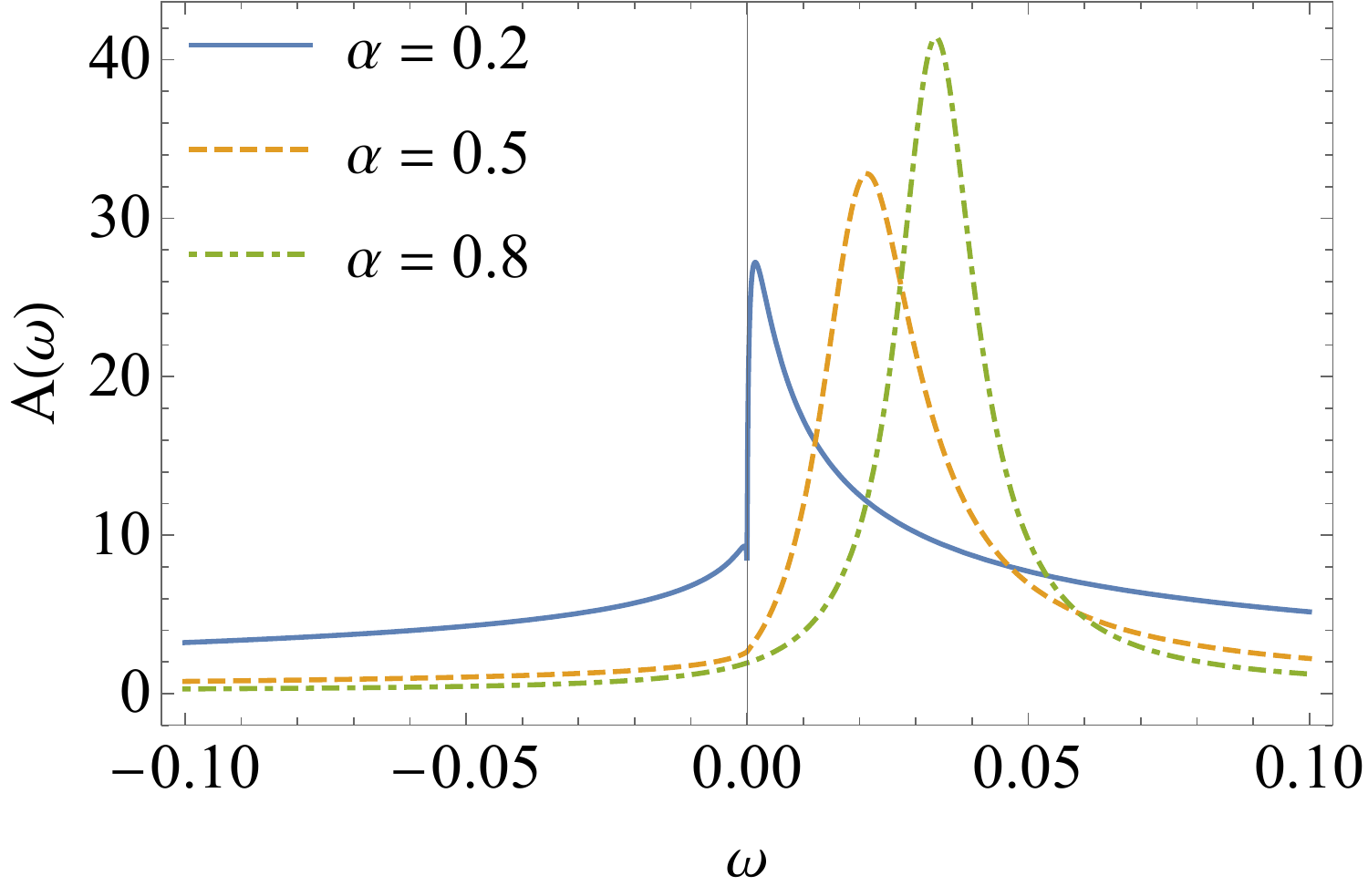}
\par\end{centering}
}

\caption{(a) The energy and momentum dependence of a power-law liquid's spectral
function for a quadratic energy spectrum with $\alpha=0.2$. (b) An
energy cut of the same plot at a momentum close to the Fermi momentum,
illustrating the shifting and broadening of the spectral function
due to the self-energy. \label{fig:spectral-function}}
\end{figure}

Next, the spectral function given by $-\im G$ is 
\begin{eqnarray}
A\left(k,\omega\right) & = & N\frac{-\Sigma^{\prpr}\left(\omega\right)}{\left[\omega-\epsilon_{k}-\Sigma^{\pr}\left(\omega\right)\right]^{2}+\left[\Sigma^{\prpr}\left(\omega\right)\right]^{2}},\label{eq:PLL-spectral-function}
\end{eqnarray}
where $N$ is a normalization constant dependent only on $\alpha$
\footnote{Due to how the self-energy $\Sigma^{\pr\pr}$ from Eq. \ref{eq:PLL-imSigma}
was obtained from ARPES momentum distribution curves, $N$ is momentum
independent.}. To make comparisons between different values of $\alpha$, we define
$N$ such that the sum rule $\int_{-\omega_{n}}^{\omega_{n}}d\omega A\left(k=k_{F},\omega\right)=1$
is obeyed for all $\alpha$'s, where $\omega_{n}$ is a high-energy
cutoff which we fix at $0.05\omega_{N}$. Fig. \ref{fig:spectral-function}
illustrates the increased shifting and broadening of the spectral
function as $\alpha$ decreases. This effect is also reflected in the density of states discussed in App. \ref{sec:dos}. Finally, since the self-energy at
the Fermi level is $\alpha$-independent, so is the Fermi momentum
$k_{F}$, and the Fermi surface remains sharp even when $\alpha\leq\frac{1}{2}$.
A sharply defined Fermi surface despite a vanishing quasiparticle
residue represents a critical Fermi surface \cite{Senthil2008}.

\section{Superconducting $T_{c}$}\label{sec:Tc}

\begin{figure}[H]
\subfloat[\label{fig:Tc-vs-alpha}]{\begin{centering}
\includegraphics[scale=0.5]{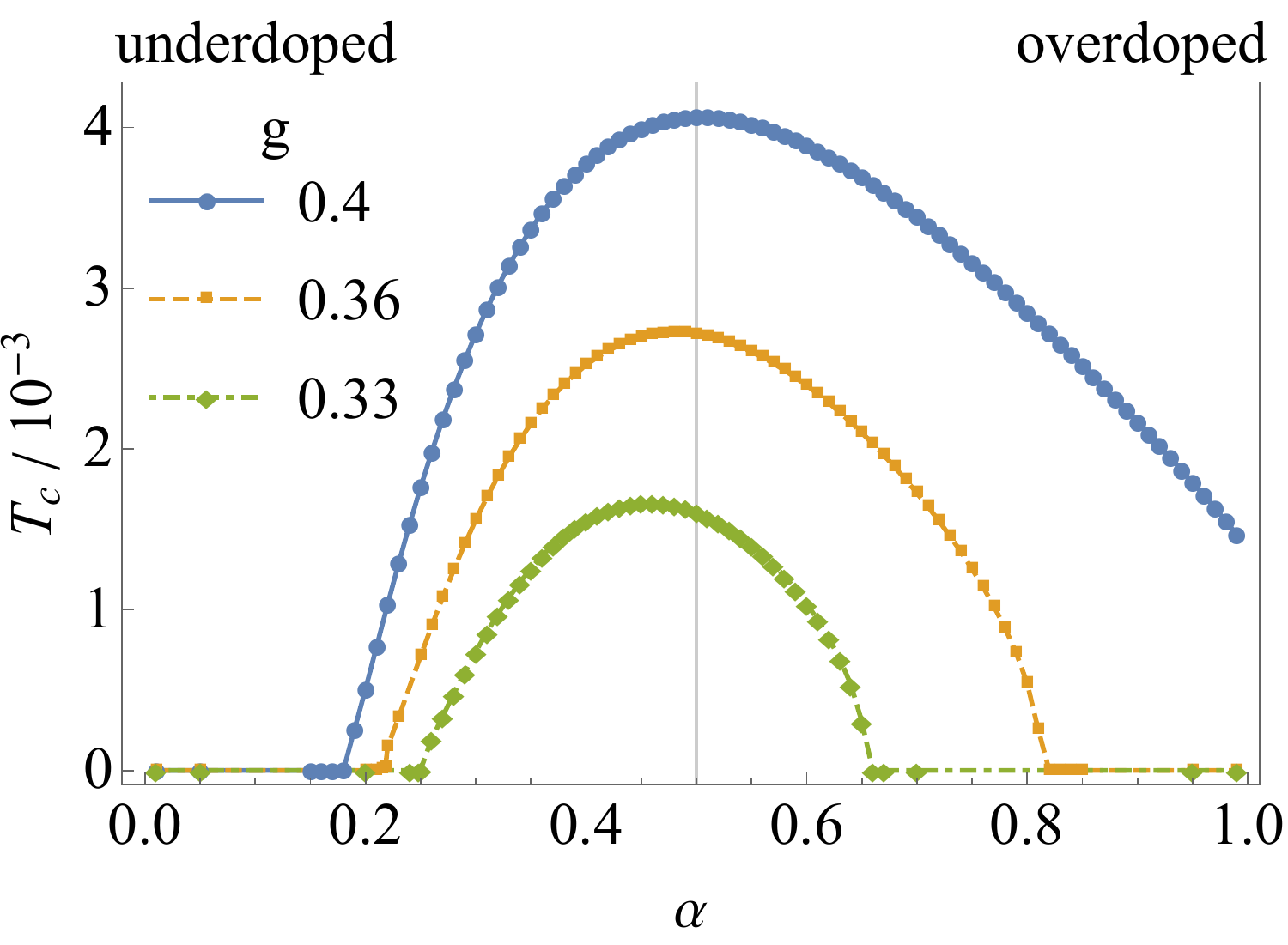}
\par\end{centering}
}

\subfloat[\label{fig:gmin-vs-alpha}]{\begin{centering}
\includegraphics[scale=0.5]{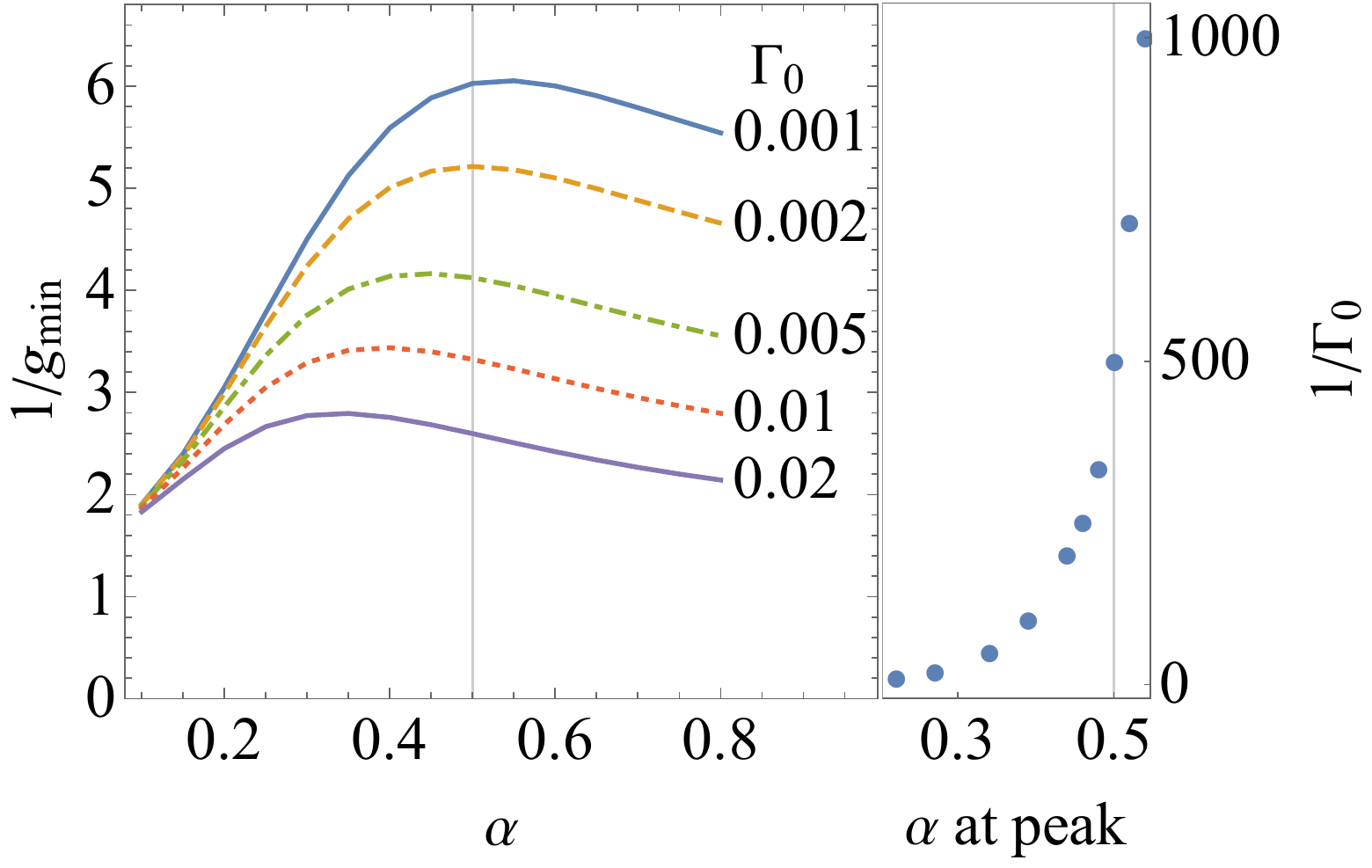}
\par\end{centering}
}

\caption{(a) The superconducting $T_{c}$ of a power-law liquid peaks at 
$\alpha=\frac{1}{2}$, reproducing the cuprates' superconducting dome.
The BCS coupling $g$ is chosen to be constant up to an energy $\omega_{D}=0.05$,
and the impurity term $\Gamma_{0}$ is fixed at $0.01$. (b) Left: The minimal
coupling $g_{\min}$ needed for superconductivity is non-monotonic
with respect to $\alpha$. Right: The peak of $1/g_{\min}$ approaches $\alpha\approx\frac{1}{2}$
as $\Gamma_{0}$ decreases. Very small values ($\lesssim 0.002$) of $\Gamma_0$ are anomalous due to numerical uncertainties.}

\end{figure}

Next, we focus on the superconducting properties of a power-law liquid.
We consider the simplest case of $s$-wave pairing symmetry with a
constant pairing interaction $g$ within an energy range $\omega_{D}$.
Within the BCS formalism, the superconducting $T_{c}$ is determined
by the pairing instability equation \cite{Phillips2013}
\begin{eqnarray}
\frac{1}{g} & = & \sum_{k}\int d\omega d\omega^{\prime}\frac{1}{2}\frac{\tanh\frac{\omega}{2T_{c}}+\tanh\frac{\omega^{\prime}}{2T_{c}}}{\omega+\omega^{\prime}}A\left(k,\omega\right)A\left(-k,\omega^{\prime}\right).\nonumber \\
\label{eq:BCS-Tc-eq}
\end{eqnarray}
From how the spectral function $A\left(k,\omega\right)$ is strongly
renormalized towards the Fermi level for $\alpha\leq\frac{1}{2}$,
we expect the superconducting $T_{c}$ to monotonically increase as
$\alpha$ decreases. However, numerical solutions to the instability
equation show that the superconducting $T_{c}$ is non-monotonic with
respect to $\alpha$, peaking  at $\alpha=\frac{1}{2}$. Shown
in Fig. \ref{fig:Tc-vs-alpha} is a power-law liquid reproducing the
cuprates' superconducting dome.  This is the central result of this paper.

To understand the origin of the $T_{c}$'s $\alpha$-dependence, we
consider the minimal BCS coupling $g_{\min}$ needed for superconductivity
by setting $T_{c}=0$ in Eq. \ref{eq:BCS-Tc-eq}:

\begin{eqnarray}
\frac{1}{g_{\min}} & = & \int d\epsilon\int d\omega d\omega^{\pr}\frac{1}{2}\frac{1}{\omega+\omega^{\pr}}A\left(\epsilon,\omega\right)A\left(\epsilon,\omega^{\pr}\right).\label{eq:BCS-gmin}
\end{eqnarray}
Fig. \ref{fig:gmin-vs-alpha} shows that $g_{\min}$ is non-monotonic
with respect to $\alpha$. In particular, the peak of $1/g_{\min}$
approaches $\alpha\approx\frac{1}{2}$ as the impurity term $\Gamma_{0}$ decreases.
A similar behavior in fact appears in Fig. \ref{fig:Tc-vs-alpha}
where the peak $T_{c}$ approaches $\alpha=\frac{1}{2}$ as $g$ increases.
When $g$ increases, superconductivity onsets in a higher temperature
regime where the impurity term $\Gamma_{0}$ is less significant.
This implies that as $\Gamma_{0}/T_{c}$ decreases, the peak $T_{c}$
approaches $\alpha=\frac{1}{2}$. These behaviors suggest a closer
study of the $\Gamma_{0}=0$ case. Since solutions for $g_{\min}$
and $T_{c}$ at $\Gamma_{0}=0$ are numerically inaccessible, we proceed
with a scaling argument.

Note that in Fig. \ref{fig:gmin-vs-alpha}, small values ($\lesssim 0.002$) of $\Gamma_0$ are anomalous because of numerical uncertainties associated with convergence issues. It is for the same reason the desired results for $\Gamma_0=0$ are numerically inaccessible and require the following scaling argument.

When $\Gamma_{0}=0$, the spectral function close to the Fermi level
($\epsilon,\omega\rightarrow0$) has the scaling form 

\begin{eqnarray}
A\left(\epsilon=r,\omega=r\right) & \sim & \frac{r^{2\alpha}}{\left(r+r^{2\alpha}\right)^{2}+r^{4\alpha}}\nonumber \\
 & \sim & \begin{cases}
r^{2\alpha-2}, & \alpha>\frac{1}{2},\\
r^{-2\alpha}, & \alpha<\frac{1}{2}.
\end{cases}
\end{eqnarray}
The two cases arise from the competition between linear and nonlinear
terms in the denominator. More concisely, 
\begin{eqnarray}
A\left(r,r\right) & \sim & r^{\xi\left(\alpha\right)},\label{eq:spectral-function-scaling}
\end{eqnarray}
with the scaling exponent
\begin{eqnarray}
\xi\left(\alpha\right) & = & 2\left|\alpha-\frac{1}{2}\right|-1. \label{eq:spectral-function-scaling-exponent}
\end{eqnarray}
This means that the spectral function's scaling exponent $\xi\left(\alpha\right)$ has a minimum
value of $-1$ at $\alpha=\frac{1}{2}$, as illustrated by the orange dashed line in Fig. \ref{fig:spectral-function-scaling-exponent}.
This result can also be verified numerically for $\Gamma_0=0$. First, the linearity of the log-log plot in the inset illustrates that the spectral function $A(k=k_F,\omega)$ from Eq. 5 indeed has a scaling form for the energy range shown. Then, the scaling exponent obtained by numerical fits is indicated by the blue solid line in the main figure. 
The solid line from numerical fits slightly differs from the analytic results because the latter is obtained in the $\omega\rightarrow 0$ limit while the former is a fit over a finite energy window. One can easily verify that the numerical results approach the analytic ones if the energy window is taken to the same limit. More precisely, the blue numerical result at $\alpha=0.5$ approaches $-1$ in the limit $\omega\rightarrow 0$, in agreement with the analytic results.

Now, consider the integral for $g_{\min}$ in spherical coordinates $\left(r,\theta,\phi\right)$
near the origin:

\begin{eqnarray}
\frac{1}{g_{\min}} & \sim & \int r^{2}dr\frac{1}{r}A\left(r,r\right)A\left(r,r\right)\nonumber \\
 & \sim & \int dr\ r^{4\left|\alpha-\frac{1}{2}\right|-1}.
\end{eqnarray}
Simply counting the powers of $r$ reveals that the integral diverges
logarithmically at $\alpha=\frac{1}{2}$ . This implies that $g_{\min}\sim0$
at $\alpha=\frac{1}{2}$, and a power-law liquid becomes most susceptible
to superconductivity. Therefore, the superconducting dome in Fig. \ref{fig:Tc-vs-alpha}
can fundamentally be attributed to the scaling form of the spectral
function.

\begin{center}
\begin{figure}[H]
\begin{centering}
\includegraphics[scale=0.5]{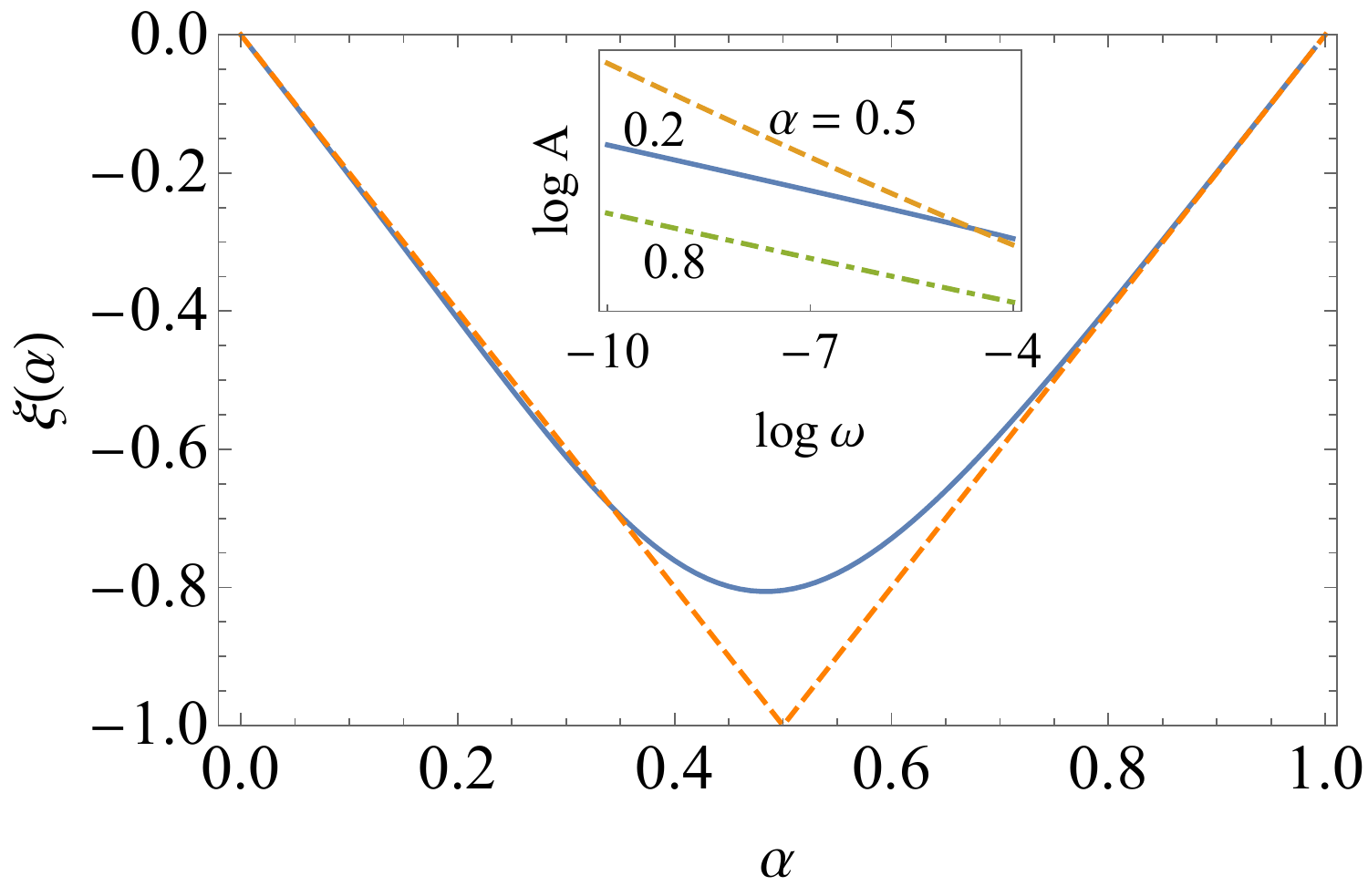}
\par\end{centering}
\caption{The scaling exponent $\xi\left(\alpha\right)$ of the spectral function $A\left(k,\omega\right)$
close to the Fermi level for $\Gamma_{0}=0$. The exponent is minimum
at $\alpha=\frac{1}{2}$. The solid line is obtained from numerical
fits over the energy range shown in the inset, while the dashed line
is based on analytic calculations from Eq. \ref{eq:spectral-function-scaling-exponent}. Inset: A log-log plot of the spectral
function close to the Fermi level for $\alpha=0.2,0.5,0.8$. \label{fig:spectral-function-scaling-exponent}}
\end{figure}
\par\end{center}

\section{Discussions}

We conclude with five pertinent points. First, we have used an $\alpha$-independent
impurity scattering term $\Gamma_{0}$ in our calculations. Experimentally,
$\Gamma_{0}$ in fact varies with $\alpha$ \cite{Reber2015}. It
is minimum ($\sim8\meV$) at optimal doping and about four (two) times
larger with underdoping (overdoping). Since impurity scattering opposes
superconductivity, one can show that such a $\Gamma_{0}$ produces
a narrower superconducting dome.

Second, let us reiterate a subtle point about Fig. \ref{fig:Tc-vs-alpha}. According to the scaling argument, the $T_c$ peaks at $\alpha=\frac{1}{2}$ because of the low-energy scaling of the spectral function when $\Gamma_0=0$. Since $\Gamma_0$ is nonzero in Fig. \ref{fig:Tc-vs-alpha}, the peak $T_c$ naturally deviates from $\alpha=\frac{1}{2}$. More precisely, notice that the deviation increases as $g$ decreases, for a fixed $\Gamma_0$. What is happening is that when $g$ decreases, superconductivity onsets in a lower temperature regime where the impurity term $\Gamma_0$ is more significant, resulting in a larger deviation from $\alpha=\frac{1}{2}$. Nevertheless, the results in Fig. 4a require $g$ to be sufficiently small. This is so that the resulting $T_c$ is low enough for the physics to be dominated by the low-energy scaling behavior of the spectral function $A$ given by Eq. \ref{eq:spectral-function-scaling}. For sufficiently large $g$, the scaling argument for the superconducting dome in the preceding section is inapplicable, and indeed we find that the $T_{c}$ becomes monotonic, and the superconducting dome vanishes.

Third, superconducting domes in other unconventional superconductors
have been attributed to various mechanisms \cite{Das2016}. In $\text{Sr}\text{Ti}\text{O}_{3}$,
screening effects \cite{Koonce1967}, longitudinal optical phonons
\cite{Gorkov2016}, and a quantum critical point \cite{Edge2015}
have been suggested. Quasiparticle-phonon interactions in dichalcogenides
\cite{Das2015,Calandra2011} and a Mott transition in organic superconductors
\cite{Hebert2015} have also been proposed. For the cuprates, self-energy
effects near the charge-density wave instability have been theorized
\cite{Buzon2014}. Our present results show that the power-law self-energy
inferred from ARPES experiments can produce the superconducting dome.

Fourth, the self-energy in Eq. \ref{eq:PLL-imSigma} was obtained
from measurements along the nodal lines of the cuprates. It is true that an accurate calculation of the superconducting $T_c$ would require measurements over the entire Brillouin zone. However, this was not our goal. Our goal was to study the implications of the novel self-energy revealed by the ARPES measurements. Given the lack of experimental data for the non-nodal regions, the most direct approach naturally assumes that the scaling form is applicable throughout the whole Brillouin zone. Doing otherwise would unnecessarily obfuscate the results which demonstrate a novel mechanism for obtaining the cuprates' superconducting dome. It is worth highlighting that, recently, similar measurements found that the antinodal self-energies are a
few times larger \cite{Li2018}. Furthermore, as the superconducting
gap develops, $\Sigma^{\prpr}$ markedly decreases while $\Sigma^{\pr}$
increases. This implies that correlations in the normal state are
converted into a strongly renormalized coherent state below $T_{c}$.
It will be interesting to incorporate these effects into the power-law
liquid model in a future work. 

Fifth, our superconducting $T_c$ calculations adopt the simplest case of $s$-wave gap symmetry, in contrast to the $d$-wave symmetry known in the cuprates. As presented in Sec. \ref{sec:Tc}, the key feature of our results arises from the scaling form of the spectral function given by Eq. \ref{eq:spectral-function-scaling}. This scaling form is intrinsic to the power-law self-energy, independent of the superconducting gap symmetry. What a $d$-wave symmetry modifies is the momentum dependence in the pairing instability equation in Eq. \ref{eq:BCS-Tc-eq}; the form of the equation's dependence on the spectral function would remain unchanged. Therefore, our results are applicable even in the $d$-wave cuprates. 

In conclusion, we studied the superconducting $T_{c}$ of a power-law
liquid, an unconventional state of matter revealed in superconducting
cuprates by recent ARPES measurements \cite{Reber2015}. The imaginary
part of the electron self-energy has the scaling form $\left(\omega^{2}+\pi^{2}T^{2}\right)^{\alpha}$,
where the scaling exponent $\alpha$ varies from $\alpha\lesssim1$
at overdoping to $\alpha\sim\frac{1}{2}$ at optimal doping, and to
$\alpha\lesssim\frac{1}{2}$ at underdoping. We found that strong
renormalization of the spectral weights results in a vanishing Fermi
velocity and diverging effective mass for $\alpha\leq\frac{1}{2}$,
in agreement with earlier experimental observations \cite{Vishik2010,Sebastian2010,Singleton2010}. 
Within a BCS formalism, we found that the superconducting
$T_{c}$ is non-monotonic with respect to $\alpha$. The $T_{c}$
peaks at around $\alpha\sim\frac{1}{2}$, reproducing the cuprates'
superconducting dome. We attribute this behavior to the low-energy
scaling form of the spectral function, where the scaling exponent
is minimum at $\alpha=\frac{1}{2}$. Our results suggest that a power-law
liquid contains physics central to understanding cuprate superconductors.
\begin{acknowledgments}
We thank the NSF DMR-1461952 for partial funding of this project.
ZL is supported by the Department of Physics at the University
of Illinois and a scholarship from the Agency of Science, Technology
and Research. CS and PWP are supported by the Center for Emergent
Superconductivity, a DOE Energy Frontier Research Center, Grant No.
DE-AC0298CH1088. KL is supported by the Department of Physics at the
University of Illinois and a scholarship from the Ministry of Science
and Technology, Royal Thai Government. 
\end{acknowledgments}

\onecolumngrid

\appendix

\section{Analytic evaluation of $\Sigma^{\protect\pr}$ \label{sec:reSigma-derivation}}

Here, we derive the real part of the self-energy, using identities
from the Digital Library of Mathematical Functions (DLMF) \cite{DLMF}.
The derivation is lengthy as a shorter one (probably using contour
integration) currently eludes us. 

From $\Sigma^{\prpr}$ in Eq. \ref{eq:PLL-imSigma}, the real part
of the self-energy via Kramers-Kronig relations (for $\left|\omega\right|<\omega_{N}$)
is 

\begin{eqnarray}
\Sigma^{\pr}\left(x\omega_{N}\right) & = & -\frac{1}{\pi}\mathcal{P}\int_{-\omega_{N}}^{\omega_{N}}\frac{d\omega^{\pr}}{\omega^{\pr}-x\omega_{N}}\left(\Gamma_{0}+\lambda\frac{\left|\omega^{\pr}\right|^{2\alpha}}{\omega_{N}^{2\alpha-1}}\right)\nonumber \\
 & = & -\frac{1}{\pi}\mathcal{P}\int_{-1}^{1}\frac{dz}{z-x}\left(\Gamma_{0}+\lambda\omega_{N}\left|z\right|^{2\alpha}\right),
\end{eqnarray}
Since we are interested only in low energy behaviors, effects from
$\left|\omega\right|>\omega_{N}$ should be negligible.

The integral over the constant impurity term is straightforward:

\begin{eqnarray}
\mathcal{P}\int_{-1}^{1}\frac{dz}{z-x} & = & -2\text{artanh}x.
\end{eqnarray}
For the second term, we break the integral into two, one with the
divergence and the other without:

\begin{eqnarray}
\mathcal{P}\int_{-1}^{1}dz\frac{\left|z\right|^{2\alpha}}{z-x} & = & \mathcal{P}\int_{0}^{1}dz\left(\frac{z^{2\alpha}}{z-x}-\frac{z^{2\alpha}}{z+x}\right)\nonumber \\
 & = & \text{sgn}\left(x\right)\left(\left|x\right|^{2\alpha}\mathcal{P}\int_{0}^{1/\left|x\right|}dz\frac{z^{2\alpha}}{z-1}-\frac{1}{\left|x\right|}\int_{0}^{1}dz\frac{z^{2\alpha}}{z/\left|x\right|+1}\right).\label{eq:cauchy-integral}
\end{eqnarray}
By series expansion and Eq. DLMF-15.8.2, the hypergeometric function
$\hf\left(a,b;c;z\right)$ has the integral representations

\begin{eqnarray}
\int dz\frac{z^{2\alpha}}{z-1} & = & -\frac{z^{1+2\alpha}}{1+2\alpha}\hf\left(1,1+2\alpha;2+2\alpha;z\right)\nonumber \\
 & = & \frac{z^{2\alpha}}{2\alpha}\hf\left(1,-2\alpha;1-2\alpha;\frac{1}{z}\right)-\pi\csc\left(2\alpha\pi\right)\left(-1\right)^{-2\alpha}.
\end{eqnarray}
These allow us to write the first integral in Eq. \ref{eq:cauchy-integral}
as 

\begin{eqnarray}
\mathcal{P}\int_{0}^{1/\left|x\right|}dz\frac{z^{2\alpha}}{z-1} & = & \left(\int_{1+\epsilon}^{1/\left|x\right|}+\int_{0}^{1-\epsilon}\right)dz\frac{z^{2\alpha}}{z-1}\nonumber \\
 & = & \frac{1}{2\alpha}\frac{1}{\left|x\right|^{2\alpha}}\hf\left(1,-2\alpha;1-2\alpha;\left|x\right|\right)-\frac{1}{2\alpha}\hf\left(1,-2\alpha;1-2\alpha;1-\epsilon\right)\nonumber \\
 &  & \quad-\frac{1}{1+2\alpha}\hf\left(1,1+2\alpha;2+2\alpha;1-\epsilon\right).\label{eq:cauchy-1a}
\end{eqnarray}
We resolve the $\epsilon\rightarrow0$ singularity by series expansion:
\begin{eqnarray}
\frac{1}{2\alpha}\hf\left(1,-2\alpha;1-2\alpha;1-\epsilon\right) & + & \frac{1}{1+2\alpha}\hf\left(1,1+2\alpha;2+2\alpha;1-\epsilon\right)\nonumber \\
 & = & \sum_{n=0}^{\infty}\left(-\frac{1}{n-2\alpha}+\frac{1}{n+1+2\alpha}\right)\left(1-\epsilon\right)^{n}\nonumber \\
 & = & \frac{1}{2\alpha}-\sum_{n=1}^{\infty}\frac{\left(1-\epsilon\right)^{n}}{n-2\alpha}+\sum_{n=1}^{\infty}\frac{\left(1-\epsilon\right)^{n-1}}{n+2\alpha}\nonumber \\
 & = & \frac{1}{2\alpha}-\sum_{n=1}^{\infty}\frac{4\alpha}{n^{2}-4\alpha^{2}}\nonumber \\
 & = & \pi\cot\left(2\pi\alpha\right),\label{eq:cauchy-1b}
\end{eqnarray}
where we have used Eq. DLMF-4.22.3 in the last line.

The second integral in Eq. \ref{eq:cauchy-integral} can be evaluated
using Eq. DLMF-15.6.1 and Eq. DLMF-15.8.2:

\begin{eqnarray}
\int_{0}^{1}dz\frac{z^{2\alpha}}{z/\left|x\right|+1} & = & \frac{1}{1+2\alpha}\hf\left(1,1+2\alpha;2+2\alpha;-\frac{1}{\left|x\right|}\right)\nonumber \\
 & = & \frac{\left|x\right|}{2\alpha}\hf\left(1,-2\alpha,1-2\alpha,-\left|x\right|\right)-\pi\csc\left(2\alpha\pi\right)\left|x\right|^{1+2\alpha}.\label{cauchy-2}
\end{eqnarray}
Finally, combining Eqs. \ref{eq:cauchy-1a}, \ref{eq:cauchy-1b},
and \ref{cauchy-2} gives 
\begin{eqnarray}
\mathcal{P}\int_{-1}^{1}dz\frac{\left|z\right|^{2\alpha}}{z-x} & = & \frac{1}{2\alpha}\text{sgn}\left(x\right)\left[\hf\left(1,-2\alpha;1-2\alpha;\left|x\right|\right)-\hf\left(1,-2\alpha,1-2\alpha,-\left|x\right|\right)\right]+\pi\text{sgn}\left(x\right)\frac{2\sin^{2}\left(\alpha\pi\right)}{2\sin\left(\alpha\pi\right)\cos\left(\alpha\pi\right)}\left|x\right|^{2\alpha}\nonumber \\
 & = & \frac{1}{2\alpha}\left[\hf\left(1,-2\alpha;1-2\alpha;x\right)-\hf\left(1,-2\alpha,1-2\alpha,-x\right)\right]+\pi\text{sgn}\left(x\right)\tan\left(\alpha\pi\right)\left|x\right|^{2\alpha}.
\end{eqnarray}
This result is nicely cast in an antisymmetric form, with the argument
of the hypergeometric function within its radius of convergence so
that the function is real.

\section{Density of states\label{sec:dos}}

\begin{figure}[H]
\begin{centering}
\includegraphics[scale=0.5]{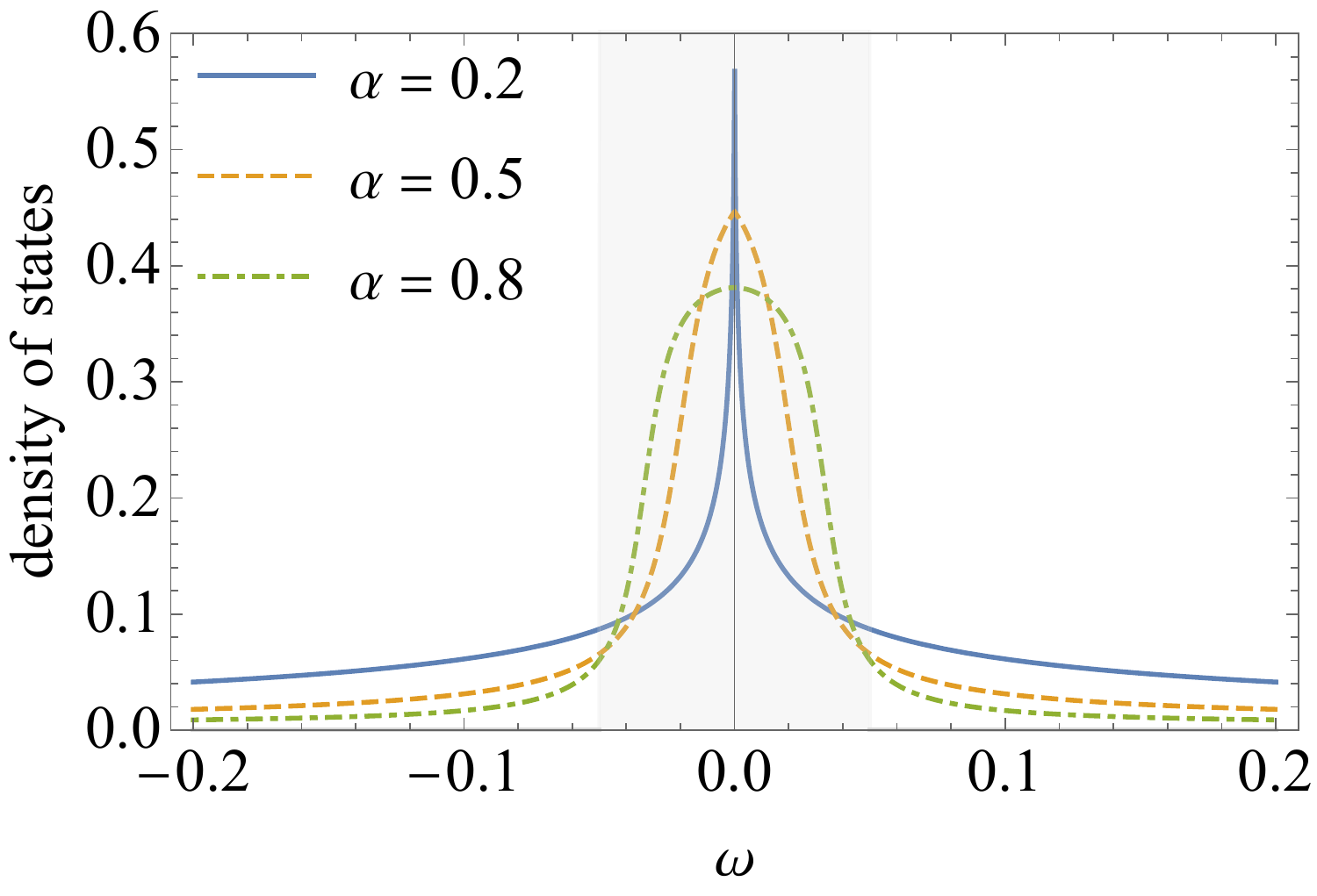}
\par\end{centering}
\caption{The density of states of a power-law liquid has a cusp at the Fermi
level for $\alpha<\frac{1}{2}$. The shaded region represents the
bare constant density of states between $\mu=\pm0.05$ for a quadratic
band in two dimensions. \label{fig:DOS-vs-omega}}
\end{figure}

In this section, we study the density of states resulting from the
shifting and broadening of the spectral function illustrated in Fig.
\ref{fig:spectral-function}. For a bare energy spectrum restricted
between $\pm\mu$, the density of states is 
\begin{eqnarray}
D\left(\omega\right) & \propto & \int\frac{d^{2}k}{\left(2\pi\right)^{2}}A\left(k,\omega\right)\nonumber \\
 & = & \frac{1}{\pi}\tan^{-1}\left[\frac{\omega+\mu-\Sigma^{\pr}\left(\omega\right)}{-\Sigma^{\pr\pr}\left(\omega\right)}\right]-\left(\mu\rightarrow-\mu\right).\label{eq:PLL-DOS}
\end{eqnarray}
Fig. \ref{fig:DOS-vs-omega} shows that the density of states greatly
deviates from a constant as $\alpha$ decreases. For $\alpha\leq\frac{1}{2}$,
it has a cusp at the Fermi level. Quantitatively, the derivative $\frac{dD}{d\omega}$
at $\omega=0$ is 
\begin{eqnarray}
\lim_{\omega\rightarrow0}\frac{dD}{d\omega} & = & \lim_{x\rightarrow0}\frac{2}{\pi\mu\omega_{N}}\frac{d\Sigma^{\pr\pr}}{dx}\nonumber \\
 & = & -\frac{4\alpha\lambda}{\pi\mu}\lim_{x\rightarrow0}\text{sgn}\left(x\right)\left|x\right|^{2\alpha-1}.\label{eq:DOS-derivative}
\end{eqnarray}
This implies that the derivative is divergent and discontinuous for
$\alpha<\frac{1}{2}$: $\lim_{\omega\rightarrow0^{\pm}}dD/d\omega=\mp\infty$.
Since this density of states is based on self-energy measured along
only the nodal lines of the cuprates, the experimental implications
of this result is unclear.

\twocolumngrid

\bibliographystyle{apsrev4-1}
\bibliography{library}

\end{document}